\documentclass[runningheads]{llncs}
\usepackage{graphicx}

    \usepackage{amsmath,amssymb}
    \usepackage{lmodern}
    \usepackage{xcolor}
    \usepackage{longtable,booktabs,array}
    \usepackage{graphicx}
    \usepackage{url}
    \usepackage{algpseudocode}
    \usepackage{algorithm}
    \usepackage{hyperref}
    \usepackage[noabbrev,capitalise]{cleveref}

    \usepackage[misc]{ifsym}
    \usepackage{orcidlink}

    \usepackage{bm}
    \usepackage{optidef}
    \usepackage{siunitx}
    \DeclareSIQualifier\dryweight{DW}
    \usepackage{chemfig}

    \usepackage[strings]{underscore}

\begin{document}
\title{LGEM\textsuperscript{+}: a first-order logic framework for automated improvement of
metabolic network models through abduction
}
\titlerunning{LGEM\textsuperscript{+}: a FOL framework for automated improvement of
MNMs}

\author{Alexander H. Gower\inst{1}\textsuperscript{(\Letter)}\orcidlink{0000-0002-8358-0842}
\and
Konstantin Korovin\inst{2}
\orcidlink{0000-0002-0740-621X}
\and
Daniel Brunnsåker\inst{1}
\orcidlink{0000-0002-5167-0536}
\and
\\
Ievgeniia A. Tiukova\inst{1,3}
\orcidlink{0000-0002-0408-3515}
\and
Ross D. King\inst{1,4,5}
\orcidlink{0000-0001-7208-4387}
}
\authorrunning{A.H. Gower et al.}
\institute{Chalmers University of Technology, Gothenburg, Sweden
\email{\{gower,danbru,tiukova,rossk\}@chalmers.se} \\
\and
The University of Manchester, Manchester, United Kingdom
\email{Konstantin.Korovin@manchester.ac.uk} \\
\and
KTH Royal Institute of Technology, Stockholm, Sweden \and
Cambridge University, Cambridge, United Kingdom \and
Alan Turing Institute, London, United Kingdom}
\maketitle        

\begin{abstract}

    Scientific discovery in biology is difficult due to the complexity of the systems involved and the expense of obtaining high quality experimental data. 
    Automated techniques are a promising way to make scientific discoveries
    at the scale and pace required to model large biological systems. 
    A key problem for 21st century biology is to build a computational model of the eukaryotic cell.
    The yeast \emph{Saccharomyces cerevisiae} is the best understood eukaryote, and genome-scale metabolic models (GEMs) are
    rich sources of background knowledge
    that we can use as a basis for automated inference and investigation.

    We present LGEM\textsuperscript{+},
    a system for automated abductive improvement of GEMs consisting of:
    a compartmentalised first-order logic framework for describing biochemical pathways
    (using curated GEMs as the expert knowledge source);
    and a two-stage hypothesis abduction procedure.
    
    We demonstrate that deductive inference on logical theories created using LGEM\textsuperscript{+}, using the
    automated theorem prover iProver, can predict
    growth/no-growth of \textit{S. cerevisiae} strains in minimal media.
    LGEM\textsuperscript{+}
    proposed 2094 unique candidate hypotheses for model improvement.
    We assess the value of the
    generated hypotheses using two criteria: (a) genome-wide single-gene
    essentiality prediction, and (b) constraint of flux-balance analysis (FBA)
    simulations. For (b) we developed an algorithm to integrate FBA with the logic model.
    We rank and filter the hypotheses using these assessments.
    We intend to test these hypotheses using the robot scientist Genesis, which is based around chemostat cultivation and high-throughput metabolomics.

\keywords{Scientific discovery \and artificial intelligence \and systems biology \and metabolic modelling \and first-order logic \and automated theorem proving.}
\end{abstract}
\section{Introduction}\label{introduction}
    
    An important aspect of modern biological science is improving our understanding of cellular processes and making sense of the complex interactions between genes, proteins and chemical species. Systems biology is the research discipline that tackles this complexity.
    \textit{Saccharomyces~cerevisiae}, commonly known as ``baker's yeast'', is an excellent model organism used for the study of eukaryote biology.
    This is due to the availability of tools for easy genetic manipulation, and low cultivation cost, which enables the design of targeted experiments to characterise the system.
    \textit{S. cerevisiae}'s was the first eukaryotic genome to be fully sequenced\cite{goffeauLife6000Genes1996_shortenedAuthors} and there is a wealth of knowledge about the gene functions, many of which are conserved or expected to have equivalents in other eukaryotes, including humans\cite{dujonYeastEvolutionaryGenomics2010}.
    Metabolic network models (MNMs) represent the cellular biochemistry of an
    organism and the related action of enzymatic genes; such models which seek to integrate knowledge from the entire organism are known as genome-scale metabolic models (GEMs).
    
    The scientific discovery problem 
    we address is
    to add knowledge to or reduce \textit{S.~cerevisiae} GEMs such that quality is increased.
    Model quality in GEMs is multi-faceted---desirable properties of a model include:
    predictive power (how well deductions using the model match experimental data); 
    network coverage (the extent to which different parts of metabolism are represented in the model); and parsimony.
    There is evidence to suggest that there are trade-offs between different desirable properties\cite{heavnerComparativeAnalysisYeast2015}.
    Foremost, however, is the predictive power of the GEM.
    Ultimately the aim is to understand the entities, mechanisms and adaptations that govern yeast growth in different environments.

    Given a draft model, improvement consists broadly of three stages: hypothesise refinements to the model; conversion of hypotheses and resultant model to a format suitable for simulation; and evaluation based on experimental evidence and internal consistency\cite{thieleProtocolGeneratingHighquality2010}. Repetition of these stages consists a scientific process which should result in a better GEM.
    Evaluation is dependent on executing simulations based on the model using a mathematical formalism,
    however optimising a model for a specific formalism is not the objective---any improvements that are made to a GEM within a certain framework should translate to improvements in the underlying knowledge.
    
    Challenges for the future of genome-scale modelling of \textit{S. cerevisiae} include:
    improving annotation;
    removing noise from low-confidence components;
    and adding reactions to eliminate so-called ``dead-end'' compounds\cite{chenGenomescaleModelingYeast2022}.
    To multiply the efforts of human researchers, several strands of research have looked at automating parts of the scientific method.
    GrowMatch was a technique developed to resolve inconsistencies between predictions and experimental observations of single-gene mutant strains of \textit{Escherichia coli}\cite{kumarGrowMatchAutomatedMethod2009}.
    Other approaches to metabolic network gap-filling have exploited answer-set programming, the most complete of which is MENECO which is designed to efficiently identify candidate additions to draft network models\cite{prigentMenecoTopologyBasedGapFilling2017_shortenedAuthors}.

    Logical inference can be applied to generate and improve metabolic models:
    induction
    allows us to generalise models from data;
    given a theory we can draw conclusions using deduction;
    and abduction
    enables us to ``fix'' the theory to be consistent with empirical data.
    In this work we use first-order logic (FOL) to simulate the metabolic network.
    This approach was first proposed in 2001\cite{reiserDevelopingLogicalModel2001_shortenedAuthors}.
    A FOL model was used to generate functional genomics hypotheses then tested by a robot scientist\cite{kingFunctionalGenomicHypothesis2004_shortenedAuthors}.
    A combination of logical induction and abduction was applied to identify inhibition in metabolic pathways due to the introduction of toxins\cite{tamaddoni-nezhadAbductionInductionLearning2005}.
    The first use of an FOL model to predict single-gene essentiality was in 2008\cite{whelanUsingLogicalModel2008}. In this study the authors constructed a set of programs in the logical programming language Prolog using the GEM iFF708\cite{forsterGenomeScaleReconstructionSaccharomyces2003} as the background knowledge source.
    Huginn is a tool that aims to integrate model revision with experimental design using abductive logic programming (ALP), and demonstrates the ability to improve metabolic models and suggest \textit{in vivo} experiments\cite{rozanskiAutomatingDevelopmentMetabolic2015}.

    A core advantage of our model---both over these previous FOL approaches that used Prolog, and over bespoke algorithmic methods such as MENECO---is that we can use powerful theorem proving software to perform deductive and abductive inference. This removes a large part of the burden of abductive algorithm design and simulation.

    Furthermore, our model is capable of deductive and abductive reasoning at scales far greater than previous FOL approaches.
    The ability to reason at scale is particularly important for the automation of scientific discovery in eukaryotic biology where the domain is complex and data are expensive to generate.
    
    In our case we use iProver\cite{korovinIProverInstantiationBasedTheorem2008},
    a theorem prover that has been used for applications including processor hardware verification\cite{khasidashviliPredicateEliminationPreprocessing2016}, and that
    has optimised solvers for the deductive and abductive reasoning tasks.
    The abstract definitions make the predicate structure a basis for abductive inference of hypotheses, and we extended iProver to include abduction inference.
    One current limitation of our FOL framework is that we do not include information on reaction stoichiometry. To integrate quantitative modelling, we also propose in this paper a method to combine flux balance analysis (FBA) and logical inference to validate metabolic pathway configurations found by the FOL model.
    We further propose a method to validate hypothesised improvements based on error reduction. 

    The main contributions of LGEM\textsuperscript{+} as presented in this paper are: (1) a compartmentalised FOL model of yeast metabolism; (2) a two-stage method for the abduction of novel hypotheses on improved models; (3) scalable methods for evaluating these models and hypotheses; and (4) an algorithm to integrate FBA with abductive reasoning.

\section{Methods}\label{methods}
    
    \subsection{The first-order logic
    framework}\label{subsec:the-first-order-logic-framework}
    
    We chose FOL as the language to express the mechanics of
    the biochemical pathways. FOL allows for a rich expression
    of knowledge about biological processes.
    Mechanisms such
    as reactions, enzyme catalysis and gene regulation can be expressed
    independently of specific genes, compounds or enzymes. 
    The method and model we design is independent of the specific network,
    meaning that although here we apply LGEM\textsuperscript{+} to \textit{S. cerevisiae},
    this modelling framework could equally well be applied to other
    organisms.
    
    We define five
    predicates in the first-order language:
    \textsf{met\textbackslash{}2},
    \textsf{gn\textbackslash{}1},
    \textsf{pro\textbackslash{}1},
    \textsf{enz\textbackslash{}1},
    and \textsf{rxn\textbackslash{}1}.
    The semantic interpretation of these predicates is outlined in \cref{table:predicates}.

    \begin{table*}
    \caption{Predicates used in the logical theory of yeast metabolism. Forward and reverse reactions are represented separately in the
    model, thus a ``positive flux'' through a reversed reaction indicates the
    reaction flux is negative.}
    \label{table:predicates}
    \centering
    \resizebox{\textwidth}{!}{
    \begin{tabular}{lll}
    \hline
    Predicate & Arguments                 & Natural language interpretation                                  \\ \hline
    \sf met\textbackslash{}2 \sf     & metabolite, compartment   & ``Metabolite X is present in cellular compartment Y.''             \\
    \sf gn\textbackslash{}1 \sf      & gene identifier           & ``Gene X is expressed.''                                           \\
    \sf pro\textbackslash{}1 \sf     & protein complex identifier         & ``Protein complex X is available (in every cellular compartment).'' \\
    \sf enz\textbackslash{}1 \sf     & enzyme category identifier & ``Enzyme category X is available.''                             \\
    \sf rxn\textbackslash{}1 \sf     & reaction                  & ``There is positive flux through reaction X.''                    \\ \hline
    \end{tabular}
    }
    \end{table*}
    
\begin{figure*}
        \centering
        \includegraphics[width=\textwidth]{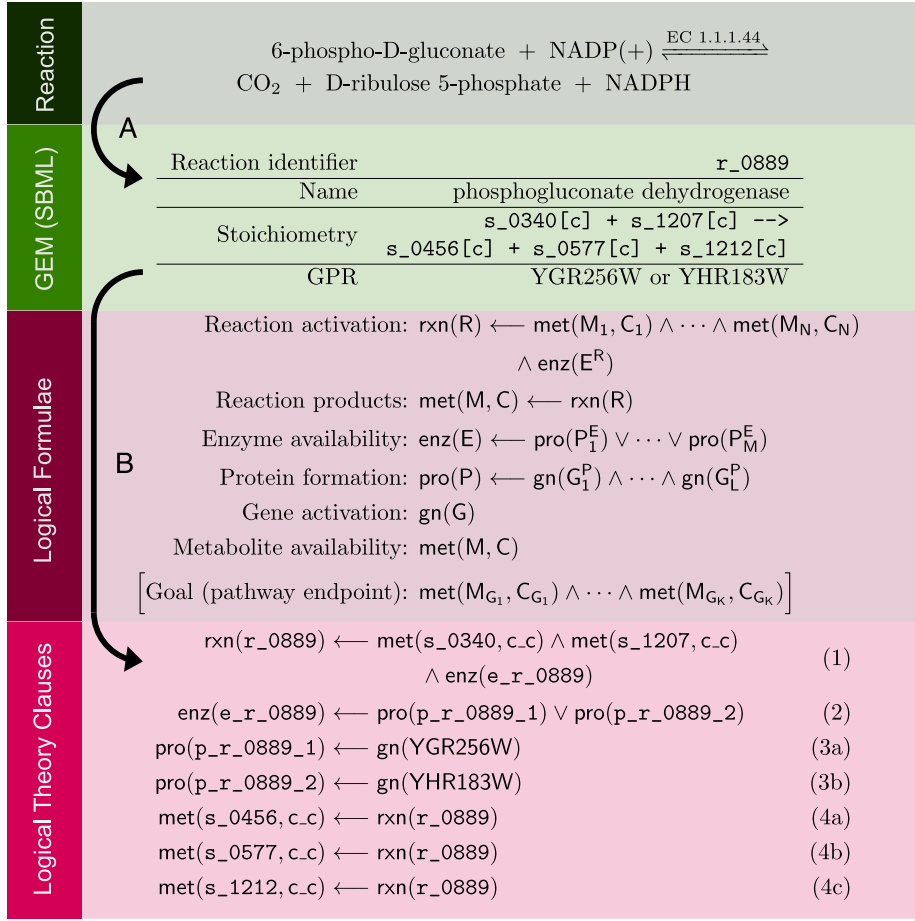}
        \caption{Conversion of genome-scale metabolic model provided in
    SBML to logical theory. \textbf{(A)} A reaction is encoded in SBML using identifiers to represent the substrates and products, and a logical rule for enzyme availability (GPR=``gene-protein-reaction rule''). \textbf{(B)} The information contained on each reaction is encoded using logical formulae into a set of clauses; predicate definitions are provided in \cref{table:predicates}.
    Here equation (1) is the reaction activation clause. ``$\land$'' is a conjunction symbol (``AND''), meaning all of the literals in the expression must be true for the RHS of the clause to be true; ``$\lor$'' is a disjunction symbol (``OR''). So we can read (1) as: ``reaction $\tt r\_0889$ is active if all of the metabolites in the set $\tt\{s\_0340, s\_1207\}$ are present in the cytoplasm and at least one of the isoenzymes is present''. Similarly equation (2) describes the condition for a relevant enzyme to be present; equations (3a,b) describe the conditions for each of these isoenzymes to be formed; and equations (4a-c) are the reaction product clauses and state that ``if reaction $\tt r\_0889$ is active then each of its products are present''.
    }
        \label{fig:gem-to-logical-model-conversion}
    \end{figure*}

    Clauses in our model are one of seven types, each expressing relationships between entities in terms of the predicates given above.
    These types of clauses are listed below, and
    we provide a graphical overview and example statements in \cref{fig:gem-to-logical-model-conversion}.
    
    \begin{itemize}
        \item \textbf{Reaction activation} clauses state that all substrate compounds for a specific reaction being present in the correct compartments, together with availability of a enzyme that can catalyse the reaction, implies the reaction is active.
        \item \textbf{Reaction product} clauses state that a reaction being active implies the presence of a product compound in a given compartment.
        \item \textbf{Enzyme availability} clauses state that the availability of the constituent parts (proteins) of an enzyme imply the availability of the enzyme. Enzymes sometimes act in complexes made up of two or more proteins, and different enzymes that catalyse the same reaction are called isoenzymes.
        \item \textbf{Protein formation} clauses state that the presence in the genome of a gene that codes for a specific protein implies the availability of that protein.
        \item \textbf{Gene presence} clauses are statements expressing either the presence or absence of a particular gene in the genome.
        \item \textbf{Metabolite presence} clauses are statements expressing the presence of a particular compound in a specific compartment.
        \item \textbf{Goal} clauses represent a biological objective, usually the presence in the cytosol of a set of compounds deemed essential for growth, but could also be another pathway endpoint or intermediary compound.
    \end{itemize}

    \subsection{Assessing growth and production of
    compounds}\label{subsec:assessing-growth-and-production-of-compounds}

    Yeast growth is dependent on the production of
    essential chemical products---intermediary points
    or endpoints of biochemical
    pathways within the organism.
    The core of these biochemical pathways is the enzymatic reactions,
    and they are facilitated by diffusion of chemicals
    within cellular compartments, including the cytosol, and passive or
    active transport across compartment boundaries or the cell membrane.
    Enzyme concentrations will depend on a wide variety of factors,
    however without the genetic code as a basis, resultant enzymes will
    not be available at all.
    Thus the genetic composition of the organism is a fundament to the
    viability of certain pathways.
    Certain products are deemed essential for growth, so if production of these compounds is inhibited then the organism is inviable.
    
    Logical inference was performed using the automated theorem proving
    software iProver (v3.7) which was chosen due to its performance and
    scalability as well as completeness for first-order theorem finding.
    The general formulation of the problem provided to iProver is to
    identify whether a theory, {\(T\)}, ``entails'' a goal, {\(G\)}.
    In other words that the goal is a logical consequence of the theory ({\(T \vDash G\)}).
    Here {\(T\)} is a set of logical axioms that encode,
    using the formalism defined in \cref{subsec:the-first-order-logic-framework}:
    knowledge from the GEM;
    the medium in which the yeast is growing,
    represented by axioms in
    the theory for the presence of compounds in the extracellular space;
    the availability of ubiquitous compounds in each cellular compartment
    and the extracellular space; and
    the presence and expression of genes.
    Deduction can be used to analyse pathways and reachable metabolites.
    In the case of growth/no-growth simulations,
    {\(G\)} represents the availability of all the essential compounds in the
    cytoplasm. So if {\(T \vDash G\)} we say that there is growth, otherwise
    not. Other goals used here are the availability of other endpoints of
    biochemical pathways. The logical proofs (that the goal is reachable) found by iProver correspond to detected biochemical pathways.

    \subsubsection{Single-gene essentiality prediction.}\label{subsec:single-gene-ess}
    
    Here we seek to predict genes without which \textit{S. cerevisiae} cannot grow.
    We compare predictions against
    lists of viable and inviable strains from a 
    genome-wide deletion mutant cultivation for \emph{S.~cerevisiae} using several media\cite{giaeverFunctionalProfilingSaccharomyces2002_shortenedAuthors}.
    In particular, we compare with cultivations on a minimal medium with the addition of uracil, histidine and leucine.
    The strain background used in this study was S288C,
    which has complete or partial deletions for HIS3, LEU2, LYS2, MET17 and URA3---for our experiments we remove these genes by default.
    Gene knockouts were performed by negating the gene presence axiom in the logical theory (i.e. $\sf gn(gene)\sf$ becomes $\sf \neg gn(gene)\sf$).

    There are two basic error types with these predictions. We follow the
    naming convention as in \cite{kumarGrowMatchAutomatedMethod2009}, that we
    have:
    (1) \emph{GNG inconsistency:} a prediction of growth when experimental
    data show no growth; and
    (2) \emph{NGG inconsistency:} a prediction of no growth when
    experimental data show growth.
    Inconsistencies arise from three main sources: deficiencies in the prior knowledge; errors in the prediction process; or conflicting empirical evidence.
    However it is the deficiencies in the prior knowledge
    that are of most interest for scientific discovery, which we explore next.

    \subsection{Abduction of hypotheses}\label{subsec:abduction-of-hypotheses}

    Abduction is used to suggest hypotheses that resolve inconsistencies between our model and empirical data. We select a reasonable set of candidate hypotheses through a two-stage process: firstly, we generate hypotheses; and secondly, we rank and filter these according to relevant scientific criteria. Generating hypotheses using an automated theorem prover is general purpose. Ranking and filtering heuristics will be domain-specific; here we describe the heuristics that we used, but others could well be applied. Pseudo-code for the abduction algorithm is provided in \cref{alg:abduction}.

    \subsubsection{Generating candidate hypotheses using iProver.}
    In the case that the goal is not reachable (i.e.
    {\(T \nvDash G\)}) iProver will suggest candidate
    hypotheses in the form of sets {\(H_{i}\)} such that
    {\(\forall i\ (T \land H_{i} \vDash G)\)}.
    This is done by reverse consequence finding ($T \land \neg G \vDash \neg H_{i}$).
    For this project we extended iProver to include these features,
    which, not being specific to biochemical reaction networks,
    could be of use for automated discovery in other scientific domains
    by constructing an appropriate FOL model.
    
    In the case of assessing growth/no-growth, the base theory as assembled from each GEM was not able to produce all essential compounds.
    Thus the first step was to abduce a set of compounds that, once available, would enable growth in the simulation for the base strain.
    Several hypotheses were suggested by iProver, we chose the one with the least additional compounds.
    
    For NGG inconsistencies there exists a set of essential metabolites not being produced that empirical data indicate will be produced given the specified genotype and conditions---in some sense the pathways in the model are incomplete.
    Hypotheses in this scenario are those that repair an incomplete pathway: additional reactions; annotation of an isoenzyme for knocked out genes; or removal of reaction annotations for knocked out genes.
    For GNG inconsistencies there is a pathway in the model that empirical data suggest should be interrupted but is not.
    In some cases we may know exactly which pathway this should be---for example empirical data displaying auxotrophy (non-production) for a specific chemical product given the genotype and conditions.
    Otherwise we know only that the production of one or more of the essential compounds should be interrupted.
    Thus hypotheses in this scenario will be those that interrupt a complete pathway: annotation of a pathway-critical reaction with a gene that is in the set of knocked out genes; removal of an isoenzyme annotation for a reaction already annotated with one of the knocked out genes; or removal of reactions.

    \subsubsection{Heuristics for ranking and filtering hypotheses.}
    The challenge addressed here is that iProver generates many hypotheses that explain the same phenomenon.
    We filter hypotheses to only include either:
    (a) addition of one or more compounds (i.e. containing only atoms using the \textsf{met} predicate) in a compartment other than the extracellular space; or
    (b) the presence of one or more particular enzyme groups for a reaction (i.e. containing only atoms using the \textsf{enz} predicate).
    The motivation for this filtering is 
    that the subsequent model improvement step (to repair the pathway) for case (a) would be to add reactions to the model that produce the hypothesised metabolites, and for case (b) to either identify an isoenzyme for hypothesised groups or remove the annotation for the deleted gene for one of these reactions.
    We also remove hypotheses that introduced availability of one or more of the target compounds in the cytosol, as this would directly ensure the goal was reached but is of no scientific value.

    We applied two criteria to assess the merit of each hypothesis. Firstly, by using the reactions activated in the proof found by iProver for each hypothesis to constrain a flux balance analysis (FBA) simulation. Around half of the hypotheses resulted in infeasible solutions or very small growth---this means perhaps there might be something else that is missing from the model, and so we have not got a reasonable hypothesis.
    The second criteria was evaluating the impact each hypothesis had on the overall error in single-gene essentiality prediction.
    If the total number of NGG errors fixed is greater than the number of GNG errors introduced then this is a good hypothesis. Another, more conservative, approach would be to only add hypotheses to the model that do not introduce any GNG errors.

    A final property that we used
    to assess the hypotheses was whether they contained compounds that were not produced by any reaction in the GEM. If this is the case, finding a suitable reaction that produces this compound and inserting into the model would repair the error.
    These hypotheses could be tested experimentally by constructing a deletion mutant, cultivating with minimal medium and after observing growth, using metabolomic analysis (e.g. with mass spectrometry) to identify if the hypothesised intermediary metabolite set is present. If so, this would provide validation for this hypothesis.
    
    If there were a reaction already in the GEM that produced the compound then it could be that there are other deficiencies in the model that need to be addressed first, for example the annotation for those reactions. In this case iProver will also find hypotheses of case (b) above. Currently we can make hypotheses to remove annotation for the knocked out gene, but this could be extended to include a search for an isoenzyme based on similarity
    (e.g. sequence similarity)
    to the knocked out gene.
    
    \begin{algorithm*}
    \caption{Abduction using logical theory of yeast metabolism}\label{alg:abduction}
    \begin{algorithmic}[1]
    \Procedure{AbductionSingleGene}{}
        \State $\mathcal{H}\gets\emptyset$
        \For{$\sf gene\sf$ in all genes in theory}
            \State $\widetilde{T}\gets T$ \Comment{Make a copy of the base theory}
            \State $\widetilde{T}\gets\widetilde{T}\setminus \{\sf gn(gene)\sf\}\cup\{\sf \neg gn(gene)\sf\}$ \Comment{Construct deletant}
            \State Use iProver to deduce if goal is reachable by identifying if $\widetilde{T}  \vDash G $
            \If{$\widetilde{T}  \vDash G $} \Comment{Growth prediction}
                \State \textbf{continue}
            \ElsIf{$\widetilde{T}  \nvDash G $} \Comment{Non-growth prediction}
                \If{$\sf gene$ is essential} \Comment{No growth observed; no error}
                    \State \textbf{continue}
                \ElsIf{$\sf gene$ is not essential} \Comment{Growth observed; NGG error}
                    \State Abduction of potential hypotheses set $\mathcal{H}_{\sf gene\sf}$ using iProver
                    \State $\mathcal{H}\gets \mathcal{H}\cup\mathcal{H}_{\sf gene\sf}$
                \EndIf
            \EndIf
        \EndFor
        \State Filter and rank $\mathcal{H}=\bigcup\limits_{\sf gene\sf\in\text{theory}} \mathcal{H}_{\sf gene}$, according to heuristics, e.g. \cref{subsec:abduction-of-hypotheses}
    \EndProcedure
    \end{algorithmic}
    \end{algorithm*}

    \subsection{Constraining flux balance analysis simulations using
    proofs}\label{constraining-flux-balance-analysis-simulations-using-proofs}
    
    Flux balance analysis (FBA) is 
    a method to find a reaction flux distribution given stoichiometric constraints from the GEM along with a biologically relevant optimisation objective, for example maximisation of biomass production\cite{orthWhatFluxBalance2010,garciasanchezComparisonAnalysisObjective2014}. FBA relies on assuming the metabolism is in steady state to find the reaction flux distribution,
    resulting in the constraint $S\bm{\nu}=\bm{0}$, where $S$ is the stoichiometric matrix for the metabolic network and $\bm{\nu}$ is the reaction flux vector. Each row in $S$ represents a compound and each column a reaction, with entries corresponding to the change in quantity of a compound resulting from the reaction in the forward direction; thus $S\in\mathbb{Z}^{m\times n}$, where $m$ is the number of compounds and $n$ is the number of reactions in the metabolic network. The stoichiometric matrix $S$ is a defined quantity and FBA aims to find a flux vector $\bm{\nu}$ according to the following regime, where $f(\nu_1,\ldots,\nu_n)$ encodes
    the biological objective.    
    
    \begin{maxi*}
    {\bm{\nu}\in\mathbb{R}^n}{f(\nu_1,\ldots,\nu_n)}{}{}
    \addConstraint{S\bm{\nu}=\bm{0}}{}{}
    \addConstraint{\nu_i^{\text{LB}}\leq\nu_i\leq\nu_i^{\text{UB}},}{}{\quad i=1,\ldots,n.}
    \end{maxi*}
    
    Whilst the stoichiometric matrix is intrinsic to the model, the upper and lower bounds for each reaction can be set to achieve relevant results. Existing methods to set these bounds include integrating experimental measurements of fluxes, or using enzyme turnover rates and availability\cite{domenzainReconstructionCatalogueGenomescale2022_shortenedAuthors}.
    We use FBA to assess the feasibility of proofs found using iProver by: first assuming that reactions that are activated in the proof are also
    activated in the FBA simulation and setting bounds for these reactions, accounting for direction; and then solving the resultant optimisation problem.
    We are able to do this neatly as both use the same GEM as the knowledge source. The procedure is outlined in \cref{alg:fba-constrain}.
    
    Flux values are measured in \unit{\milli\mole\per\gram\dryweight\per\hour} and metabolite concentrations vary substantially between compounds, so finding a forcing threshold which is appropriate for all reactions is not straightforward. For our FBA simulations we used the Python package \texttt{cobrapy} (version 0.26.3)\cite{ebrahimCOBRApyCOnstraintsBasedReconstruction2013};
    in the absence of relevant documentation on a suitable threshold,
    we found in a discussion for a MATLAB implementation of COBRA that a suitable threshold should be set at \num{1e-9}\cite{WhatMinimumFlux}.

    \begin{algorithm*}
    \caption{Constraining FBA solution given a logical theory $T$ and a goal $G$}\label{alg:fba-constrain}
    \begin{algorithmic}[1]
    \Function{FBAConstrain}{GEM, $T$, $G$, $\nu_0$} \Comment{$\nu_0$ is minimum flux threshold for activation}
        \State Use iProver to find proof of $T \vDash G$ \Comment{The goal is reachable}
        \State $i\gets 1$
        \While{$i\leq N$} \Comment{$N$ is the number of reactions in the GEM}
            \If {$r_i$ active in the proof in the forward direction}
                \State $\nu^{LB}_i\gets\nu_0$ \Comment{Force reactions to have positive flux}
            \ElsIf {$r_i$ active in the proof in the reverse direction}
                \State $\nu^{UB}_i\gets-\nu_0$
            \EndIf
            \State $i\gets i+1$
        \EndWhile
        \State Solve FBA problem ($S\bm{\nu}=\bm{0}$) with resultant flux bounds
        \State \textbf{return} $(\bm{\nu},\rm growthValue, solutionStatus\rm)\in\mathbb{R}^N\times\mathbb{R}\times\{\rm optimal, infeasible\rm\}$
    \EndFunction
    \end{algorithmic}
    \end{algorithm*}

\subsection{Sources of knowledge}\label{subsec:sources-of-knowledge}
    
    The primary source of the knowledge about reactions and associated genes
    is the GEM Yeast8 (v8.46.4.46.2)\cite{luConsensusCerevisiaeMetabolic2019_shortenedAuthors}. This
    was chosen due to its broad coverage of the reactions and gene
    associations as well as its specificity to the organism \emph{S.~cerevisiae}.
    The other two GEMs used were: 
    iMM904\cite{moConnectingExtracellularMetabolomic2009} and
    iFF708\cite{forsterGenomeScaleReconstructionSaccharomyces2003}.
    (We include iFF708 as a background knowledge source partly to enable comparison with previous logical modelling approach\cite{whelanUsingLogicalModel2008}.)
    The models are stored using Systems Biology Markup Language
    (SBML). The software written to convert a GEM SBML file to a logical
    knowledge base is available in the supporting material, and follows the
    process described below and shown in \cref{fig:gem-to-logical-model-conversion}.
    
    \subsection{Compiling lists of compounds from each
    model}\label{subsec:compiling-lists-of-compounds-from-each-model}
    
    We use three reference lists from \cite{whelanUsingLogicalModel2008};
    these are shown in the first column of
    the files on the LGEM\textsuperscript{+} GitHub repository\footnote{\url{https://github.com/AlecGower/LGEMPlus}} corresponding to:
    (1) all compounds deemed essential for growth in \emph{S.~cerevisiae}\footnote{\texttt{src/model-files/essential-compounds-\{model\}.tsv}};
    (2) compounds assumed ubiquitous during growth assumed to be present
    throughout the cell regardless of initial conditions, such as \chemfig{H_2O} and \chemfig{O_2}\footnote{\texttt{src/model-files/ubiquitous-compounds-\{model\}.tsv}}; and
    (3) the growth media for the experiments,
    in this case yeast nitrogen base (YNB) with addition of ammonium, glucose and three amino acids (uracil, histidine and leucine)\footnote{\texttt{src/model-files/ynb-compounds-\{model\}.tsv}}.
    
    Each compound in these lists has an associated Kyoto Encyclopedia of
    Genes and Genomes (KEGG)\cite{kanehisaKEGGKyotoEncyclopedia2000} identifier.
    Matching compounds in the curated GEMs based firstly on KEGG ID, otherwise using the species name or synonyms.
    Some of the compounds we wish to include do not have corresponding entities in the GEMs used as background knowledge.
    Therefore there are discrepancies between the reference lists and the compiled
    lists.

\section{Results}\label{results}
    
    \subsubsection{Automated theorem proving software can be used
    to estimate single-gene essentiality given a prior network
    model.}\label{automated-theorem-proving-software-iprover-can-be-used-to-estimate-single-gene-essentiality-given-a-prior-network-model.}

    Using three GEMs---Yeast8, iMM904 and iFF708---as background knowledge sources we conducted
    single-gene deletant simulations to assess essentiality of each gene and compared against a genome-wide deletion mutant cultivation\cite{giaeverFunctionalProfilingSaccharomyces2002_shortenedAuthors}.
    Detailed descriptions of these methods are provided in \cref{methods}.
    A summary of the single-gene essentiality prediction results is provided in \cref{table-performance-comparison}.
    
    In general, 
    our approach offers state-of-the-art error prediction results for a qualitative model\cite{wunderlichUsingTopologyMetabolic2006,whelanUsingLogicalModel2008}. These error rates indicate how much is still to be learnt about yeast metabolism.
    We also found that gene essentiality predictions vary somewhat depending on
    the prior.

    \begin{table*}
    \centering
    \resizebox{\textwidth}{!}{%
    \begin{tabular}{llll|ll}
    \hline
    Base GEM                               & Yeast8          & iMM904         & iFF708         & Yeast8 (FBA)   & \emph{Syn. Acc.}\cite{wunderlichUsingTopologyMetabolic2006} \\ \hline
    \# predictions (\#genes in GEM) & 1056 (1150)     & 827 (905)      & 566 (619)      & 1068 (1150)    & \emph{682}            \\
    NG Recall (NGNG/*NG)                    & 0.193 (31/161)  & 0.266 (33/124) & 0.140 (14/100) & 0.447 (72/161) & \emph{0.119 (14/118)}       \\
    NG Precision  (NGNG/NG*)         & 0.431 (31/72)   & 0.478 (33/69)  & 0.778 (14/18)  & 0.459 (72/157) & \emph{0.292 (14/48)}        \\
    GNG Rate (GNG/*NG)                      & 0.807 (130/161) & 0.734 (91/124) & 0.860 (86/100) & 0.553 (89/161) & \emph{0.881 (104/118)}      \\
    NGG Rate (NGG/*G)                       & 0.046 (41/895)  & 0.051 (36/703) & 0.009 (4/466)  & 0.094 (85/907) & \emph{0.060 (34/564)}     \\
    F1 score                               & 0.266           & 0.342          & 0.237          & 0.453          & \emph{0.169}         \\ \hline
    \end{tabular}
    }
    \caption{Comparative prediction results for single-gene essentiality using logical theory and deduction in iProver across three background knowledge sources: Yeast8 (v8.46.4.46.2); iMM904; and iFF708, with comparison to: (a) an FBA-simulation with a viability threshold on growth rate set at
    \num{1e-6} \unit{\per\hour} (according to \cite{luConsensusCerevisiaeMetabolic2019_shortenedAuthors});
    and (b) another qualitative prediction method, the ``synthetic accessibility'' approach taken by Wunderlich et. al.\cite{wunderlichUsingTopologyMetabolic2006}. The empirical data used as truth data for these statistics were taken from a genome-wide screening study using a minimal medium\cite{giaeverFunctionalProfilingSaccharomyces2002_shortenedAuthors}. The FOL model performance represents an improvement on previous qualitative method. 
    \\
    \textit{Shorthand}: *NG-observed no growth; *G-observed growth; NG*-predicted no growth. 
    (Note that the performance statistics for the synthetic accessibility method are taken directly from the authors' report so there may be a difference in truth data to those used to evaluate our model.)}
    \label{table-performance-comparison}
    \end{table*}
    
    \subsubsection{Abductive reasoning allows for identification of possible
    missing reactions.}\label{abductive-reasoning}
    
    None of the logical theories resultant from the conversion from Yeast8, iMM904 and iFF708 were viable given the minimal medium and ubiquitous compounds, even without any gene deletions, meaning one or more of the essential compounds 
    was not produced.
    iProver abduced hypotheses consisting of combinations of compounds whose presence would enable viability of the base strain.
    These abductions represent gaps in the model, possibly from missing
    reactions.

    We then apply the LGEM\textsuperscript{+} abduction procedure to model improvement, here demonstrated on the Yeast8 model.
    For each of the 41 NGG errors in the single-gene deletion task, we
    generated candidate hypotheses according to methods described in \cref{subsec:abduction-of-hypotheses}.
    In total we generated 2094 unique hypotheses; some hypotheses would result in an error correction for several genes. 
    We ranked and filtered these hypotheses according to domain-specific heuristics,
    finding 681 of these were valid,
    i.e. only containing \textsf{met} (633) or \textsf{enz} (48) predicates.
    The FBA evaluation indicated 534 hypotheses that could be balanced by the reactions forced in the model, 118 of which were valid.
    There were 14 hypotheses that were valid and also resulted in a net improvement on the single-gene prediction task. Several of these hypothesised the availability of compounds in the L-arginine biosynthesis pathway which we discuss below.

    \subsubsection{Strict essentiality criteria may
    explain NGG inconsistencies.}\label{our-criteria-for-essentiality-are-strict-which-may-explain-false-positives}

    The criteria for growth are the production of all essential metabolites, meaning if just one is not produced we have no growth.
    One result of this setup is a relatively low precision in the single-gene essentiality prediction. Of the 72 deletions predicted inviable by our model, 41 of these are shown to result in experimentally viable mutant strains (NGG errors).

    For several genes in the L-arginine biosynthesis pathway the only essential metabolite not reachable in the model was L-arginine. These resulted in NGG errors despite the pathway structure and previous empirical evidence showing that null mutants
    for \textit{ARG1}\cite{crabeelArginineRepressionSaccharomyces1988}, \textit{ARG3}\cite{chengImportProcessingHuman1987}, \textit{ARG4}\cite{hsiaoHighfrequencyTransformationYeast1979} and \textit{ARG5,6}\cite{boonchirdCharacterizationYeastARG51991}
    are auxotrophic for L-arginine (i.e. L-arginine was not produced).
    These results demonstrate that 
    the model can successfully identify behaviour of the metabolic network consistent with other experimental evidence and not the genome-wide screen results\cite{giaeverFunctionalProfilingSaccharomyces2002_shortenedAuthors}.
    These cases are candidates for experimental testing, and highlight the potential of such models to inform laboratory experimental design and research direction.

    \subsubsection{GNG errors are possibly due to shortcuts in the model
    due to incomplete annotation.}\label{false-negatives-are-likely-due-to-shortcuts-in-the-model-due-to-poor-annotation.}
    
    Given the strict essentiality criteria we would also expect this method to show high recall on (sensitivity to) inviable mutant strains. However there are still a large number of GNG errors: 130 for the logical model predictions based on Yeast8.
    Due to the method of deduction and the assumption that all genes are expressed unless they are removed, a reaction within the network is active unless 
    a gene is knocked out that is
    an essential component in each isoenzyme for that reaction.
    A reaction not annotated with an enzyme, or one annotated with a set of isoenzymes that do not have a specific gene in common, will always be activated in our model provided the required substrates are present. 
    In the Yeast8 model there are 4058 reactions, 1425 (35\%) of which have no enzyme annotation and 540 (13\%) of which satisfy the second condition. Thus nearly half of all reactions will not be affected by single-gene deletions.

\section{Discussion and Conclusion}\label{discussion-and-conclusion}
    
    Scientific discovery in biology is difficult due to the complexity of the systems involved and the expense of obtaining high quality experimental data. 
    Automated techniques that make good use of background knowledge,
    of which GEMs are prime examples,
    will have a strong starting point.
    LGEM\textsuperscript{+} seeks to do just that by using FOL combined with a powerful theorem prover, iProver.
    
    We efficiently predicted single-gene essentiality in \textit{S. cerevisiae}
    using a first-order logic (FOL) model.
    Our method showed state of the art results compared to previous
    qualitative methods, yet quantitative
    prediction using FBA achieves a higher precision and
    recall.
    
    We designed and implemented an algorithm for the abduction of hypotheses for
    improvement of a GEM.
    We found 633 hypotheses proposing availability of compounds in specific compartments,
    and therefore indicate possible missing reactions,
    118 of which were validated through FBA constraint and
    14 of which resulted in improvements in the single-gene essentiality prediction task, 
    We intend to test these hypotheses using the robot scientist Genesis, which is based around chemostat cultivation and high-throughput metabolomics.

    Measuring performance statistics relative to the number of genes in a model, rather than the number of genes in the organism, presents some challenges when designing a learning process to improve this performance (such as GrowMatch\cite{kumarGrowMatchAutomatedMethod2009}).
    This highlights the need for better model assessment criteria to drive abduction. We have attempted here to provide an example with the constraint of FBA solutions. Future work could certainly be directed to defining such criteria and integrating them into LGEM\textsuperscript{+}.
    
    The logical theory developed here was focused on efficient inference on biochemical pathways. A challenge for future development is to extend this theory, for example more detailed encoding of enzyme availability or integration of gene regulation and signalling.
    Aligning the logic more closely with existing ontologies, for example the Systems Biology Ontology (SBO),
    would ensure the theory remains useful and semantically precise as it is extended. This is a common challenge across the scientific discovery community as we move further toward joint teams of human and robot scientists---ontologies provide a common language.
    
    The best way to test hypotheses is through \textit{in vivo} experimentation. Integrating LGEM\textsuperscript{+} into an automated experimental design process would enable the next generation of robot scientists.

\subsubsection{Code and Data Availability:}
Code and data used in this study, including the tables for essential compounds, ubiquitous compounds and minimal media, are available at \url{https://github.com/AlecGower/LGEMPlus}.

\textbf{Acknowledgements:}
This work was partially supported by the Wallenberg AI, Autonomous Systems and Software Program (WASP) funded by the Alice Wallenberg Foundation. Funding was also provided by the Chalmers AI Research Centre and the UK Engineering and Physical Sciences Research Council (EPSRC) grant nos: EP/R022925/2 and EP/W004801/1, as well as the Swedish
Research Council Formas (2020-01690).

\bibliographystyle{splncs04}
\bibliography{references}

\begin{thebibliography}{10}
\providecommand{\url}[1]{\texttt{#1}}
\providecommand{\urlprefix}{URL }
\providecommand{\doi}[1]{https://doi.org/#1}

\bibitem{boonchirdCharacterizationYeastARG51991}
Boonchird, C., Messenguy, F., Dubois, E.: Characterization of the yeast
  {{ARG5}},6 gene: Determination of the nucleotide sequence, analysis of the
  control region and of {{ARG5}},6 transcript. Molecular \& general genetics:
  MGG  \textbf{226}(1-2),  154--166 (Apr 1991). \doi{10.1007/BF00273599}

\bibitem{chenGenomescaleModelingYeast2022}
Chen, Y., Li, F., Nielsen, J.: Genome-scale modeling of yeast metabolism:
  Retrospectives and perspectives. FEMS Yeast Research  \textbf{22}(1),
  foac003 (Jan 2022). \doi{10.1093/femsyr/foac003}

\bibitem{chengImportProcessingHuman1987}
Cheng, M.Y., Pollock, R.A., Hendrick, J.P., Horwich, A.L.: Import and
  processing of human ornithine transcarbamoylase precursor by mitochondria
  from {{Saccharomyces}} cerevisiae. Proceedings of the National Academy of
  Sciences of the United States of America  \textbf{84}(12),  4063--4067 (Jun
  1987). \doi{10.1073/pnas.84.12.4063}

\bibitem{WhatMinimumFlux}
Cobra-Toolbox: What is the minimum flux computed by {{Flux Balance Analysis}}
  or the accuracy of {{FBA}}?
  https://groups.google.com/g/cobra-toolbox/c/9xmP1VcrWL0

\bibitem{crabeelArginineRepressionSaccharomyces1988}
Crabeel, M., Seneca, S., Devos, K., Glansdorff, N.: Arginine repression of the
  {{Saccharomyces}} cerevisiae {{ARG1}} gene. {{Comparison}} of the {{ARG1}}
  and {{ARG3}} control regions. Current Genetics  \textbf{13}(2),  113--124
  (Feb 1988). \doi{10.1007/BF00365645}

\bibitem{domenzainReconstructionCatalogueGenomescale2022_shortenedAuthors}
Domenzain, I., S{\'a}nchez, B., Anton, M., et~al.: Reconstruction of a
  catalogue of genome-scale metabolic models with enzymatic constraints using
  {{GECKO}} 2.0. Nature Communications  \textbf{13}(1), ~3766 (Jun 2022).
  \doi{10.1038/s41467-022-31421-1}

\bibitem{dujonYeastEvolutionaryGenomics2010}
Dujon, B.: Yeast evolutionary genomics. Nature Reviews Genetics
  \textbf{11}(7),  512--524 (Jul 2010). \doi{10.1038/nrg2811}

\bibitem{ebrahimCOBRApyCOnstraintsBasedReconstruction2013}
Ebrahim, A., Lerman, J.A., Palsson, B.O., Hyduke, D.R.: {{COBRApy}}:
  {{COnstraints-Based Reconstruction}} and {{Analysis}} for {{Python}}. BMC
  Systems Biology  \textbf{7}(1), ~74 (Aug 2013). \doi{10.1186/1752-0509-7-74}

\bibitem{forsterGenomeScaleReconstructionSaccharomyces2003}
F{\"o}rster, J., Famili, I., Fu, P., Palsson, B.{\O}., Nielsen, J.:
  Genome-{{Scale Reconstruction}} of the {{Saccharomyces}} cerevisiae
  {{Metabolic Network}}. Genome Research  \textbf{13}(2),  244--253 (Jan 2003).
  \doi{10.1101/gr.234503}

\bibitem{garciasanchezComparisonAnalysisObjective2014}
Garc{\'i}a~S{\'a}nchez, C.E., Torres~S{\'a}ez, R.G.: Comparison and analysis of
  objective functions in flux balance analysis. Biotechnology Progress
  \textbf{30}(5),  985--991 (2014). \doi{10.1002/btpr.1949}

\bibitem{giaeverFunctionalProfilingSaccharomyces2002_shortenedAuthors}
Giaever, G., Chu, A.M., Ni, L., et~al.: Functional profiling of the
  {{Saccharomyces}} cerevisiae genome. Nature  \textbf{418}(6896),  387--391
  (Jul 2002). \doi{10.1038/nature00935}

\bibitem{goffeauLife6000Genes1996_shortenedAuthors}
Goffeau, A., Barrell, B.G., Bussey, H., et~al.: Life with 6000 {{Genes}}.
  Science  \textbf{274}(5287),  546--567 (Oct 1996).
  \doi{10.1126/science.274.5287.546}

\bibitem{heavnerComparativeAnalysisYeast2015}
Heavner, B.D., Price, N.D.: Comparative {{Analysis}} of {{Yeast Metabolic
  Network Models Highlights Progress}}, {{Opportunities}} for {{Metabolic
  Reconstruction}}. PLOS Computational Biology  \textbf{11}(11),  e1004530 (Nov
  2015). \doi{10.1371/journal.pcbi.1004530}

\bibitem{hsiaoHighfrequencyTransformationYeast1979}
Hsiao, C.L., Carbon, J.: High-frequency transformation of yeast by plasmids
  containing the cloned yeast {{ARG4}} gene. Proceedings of the National
  Academy of Sciences of the United States of America  \textbf{76}(8),
  3829--3833 (Aug 1979). \doi{10.1073/pnas.76.8.3829}

\bibitem{kanehisaKEGGKyotoEncyclopedia2000}
Kanehisa, M.: {{KEGG}}: {{Kyoto Encyclopedia}} of {{Genes}} and {{Genomes}}.
  Nucleic Acids Research  \textbf{28}(1),  27--30 (Jan 2000).
  \doi{10.1093/nar/28.1.27}

\bibitem{khasidashviliPredicateEliminationPreprocessing2016}
Khasidashvili, Z., Korovin, K.: Predicate {{Elimination}} for {{Preprocessing}}
  in {{First-Order Theorem Proving}}. In: Creignou, N., Le~Berre, D. (eds.)
  Theory and {{Applications}} of {{Satisfiability Testing}} \textendash{}
  {{SAT}} 2016. pp. 361--372. Lecture {{Notes}} in {{Computer Science}},
  {Springer International Publishing}, {Cham} (2016).
  \doi{10.1007/978-3-319-40970-2_22}

\bibitem{kingFunctionalGenomicHypothesis2004_shortenedAuthors}
King, R.D., Whelan, K.E., Jones, F.M., et~al.: Functional genomic hypothesis
  generation and experimentation by a robot scientist. Nature
  \textbf{427}(6971),  247--252 (Jan 2004). \doi{10.1038/nature02236}

\bibitem{korovinIProverInstantiationBasedTheorem2008}
Korovin, K.: {{iProver}} \textendash{} {{An Instantiation-Based Theorem
  Prover}} for {{First-Order Logic}} ({{System Description}}). In: Armando, A.,
  Baumgartner, P., Dowek, G. (eds.) Automated {{Reasoning}}, vol.~5195, pp.
  292--298. {Springer Berlin Heidelberg}, {Berlin, Heidelberg} (2008).
  \doi{10.1007/978-3-540-71070-7_24}

\bibitem{kumarGrowMatchAutomatedMethod2009}
Kumar, V.S., Maranas, C.D.: {{GrowMatch}}: {{An Automated Method}} for
  {{Reconciling In Silico}}/{{In Vivo Growth Predictions}}. PLoS Computational
  Biology  \textbf{5}(3),  e1000308 (Mar 2009).
  \doi{10.1371/journal.pcbi.1000308}

\bibitem{luConsensusCerevisiaeMetabolic2019_shortenedAuthors}
Lu, H., Li, F., S{\'a}nchez, B.J., et~al.: A consensus {{S}}. cerevisiae
  metabolic model {{Yeast8}} and its ecosystem for comprehensively probing
  cellular metabolism. Nature Communications  \textbf{10}(1) (2019).
  \doi{10.1038/s41467-019-11581-3}

\bibitem{moConnectingExtracellularMetabolomic2009}
Mo, M.L., Palsson, B., Herrg{\aa}rd, M.J.: Connecting extracellular metabolomic
  measurements to intracellular flux states in yeast. BMC Systems Biology
  \textbf{3} (2009). \doi{10.1186/1752-0509-3-37}

\bibitem{orthWhatFluxBalance2010}
Orth, J.D., Thiele, I., Palsson, B.{\O}.: What is flux balance analysis? Nature
  Biotechnology  \textbf{28}(3),  245--248 (Mar 2010). \doi{10.1038/nbt.1614}

\bibitem{prigentMenecoTopologyBasedGapFilling2017_shortenedAuthors}
Prigent, S., Frioux, C., Dittami, S.M., et~al.: Meneco, a {{Topology-Based
  Gap-Filling Tool Applicable}} to {{Degraded Genome-Wide Metabolic Networks}}.
  PLOS Computational Biology  \textbf{13}(1),  e1005276 (Jan 2017).
  \doi{10.1371/journal.pcbi.1005276}

\bibitem{reiserDevelopingLogicalModel2001_shortenedAuthors}
Reiser, P.G.K., King, R.D., Muggleton, S.H.o.: Developing a logical model of
  yeast metabolism. Electronic Transactions in Artificial Intelligence
  \textbf{5}(B),  223--244 (2001)

\bibitem{rozanskiAutomatingDevelopmentMetabolic2015}
Rozanski, R., Bragaglia, S., Ray, O., King, R.: Automating the {{Development}}
  of {{Metabolic Network Models}}. In: Roux, O., Bourdon, J. (eds.)
  Computational {{Methods}} in {{Systems Biology}}. pp. 145--156. Lecture
  {{Notes}} in {{Computer Science}}, {Springer International Publishing},
  {Cham} (2015). \doi{10.1007/978-3-319-23401-4_13}

\bibitem{tamaddoni-nezhadAbductionInductionLearning2005}
{Tamaddoni-Nezhad}, A., Chaleil, R., Kakas, A., Muggleton, S.: Abduction and
  induction for learning models of inhibition in metabolic networks. In: Fourth
  {{International Conference}} on {{Machine Learning}} and {{Applications}}
  ({{ICMLA}}'05). pp. 6 pp.-- (Dec 2005). \doi{10.1109/ICMLA.2005.6}

\bibitem{thieleProtocolGeneratingHighquality2010}
Thiele, I., Palsson, B.{\O}.: A protocol for generating a high-quality
  genome-scale metabolic reconstruction. Nature Protocols  \textbf{5}(1),
  93--121 (Jan 2010). \doi{10.1038/nprot.2009.203}

\bibitem{whelanUsingLogicalModel2008}
Whelan, K.E., King, R.D.: Using a logical model to predict the growth of yeast.
  BMC bioinformatics  \textbf{9}, ~97 (Feb 2008). \doi{10.1186/1471-2105-9-97}

\bibitem{wunderlichUsingTopologyMetabolic2006}
Wunderlich, Z., Mirny, L.A.: Using the {{Topology}} of {{Metabolic Networks}}
  to {{Predict Viability}} of {{Mutant Strains}}. Biophysical Journal
  \textbf{91}(6),  2304--2311 (Sep 2006). \doi{10.1529/biophysj.105.080572}

\end{thebibliography}

\end{document}